\documentstyle[aps,graphicx]{revtex}
\begin{document}
\title{Bose-Einstein condensation of two interacting particles}

\author{Markus A. Cirone $^{1,2,3}$
\footnote[5]{Present address: Abteilung f\"ur Quantenphysik,
Universit\"at Ulm, D-89069 Ulm, Germany}
,Krzysztof G\'{o}ral $^1$,
Kazimierz Rz\c{a}\.{z}ewski $^1$
and Martin Wilkens $^4$}

\address{$^1$ Center for Theoretical Physics and
College of Science,
\\ Polish Academy of Sciences, Aleja Lotnik\'ow
32/46, 02-668 Warsaw, Poland}
\address{$^2$ Dipartimento di
Scienze Fisiche ed Astronomiche,
\\ Universit\`{a} di Palermo, via
Archirafi 36, I-90123 Palermo, Italy}
\address{$^3$ INFM and
Dipartimento di Fisica e Tecnologie Relative,
\\ Universit\`{a} di
Palermo, Viale delle Scienze, I-90128 Palermo, Italy}
\address{$^4$ Institute of Physics, University of Potsdam,
\\ Am Neuen Palais 10, D-14469 Potsdam, Germany}

\maketitle

\begin{abstract}
We investigate the notion of Bose-Einstein condensation of
interacting particles. The definition of the condensate is based
on the existence of the dominant eigenvalue of the single-particle
density matrix. The statistical properties and the characteristic
temperature are computed exactly in the soluble models of two
interacting atoms.
\end{abstract}

\section{Introduction}

The recent observation of the quantum degeneracy in mesoscopic
samples of dilute alkaline, hydrogen and metastable helium gases
\cite{becexp} marks an important breakthrough in a
long standing quest for the
experimental realization of the Bose-Einstein condensation in its
original form \cite{Bose,BEC}. Compared to other systems -- cold
gases, say, or liquid helium -- these new systems are
characterized by an unprecedented range in life-time, density, and
temperature, by their purity and variability, and by the precision
with which they can be manipulated and controlled. The precision
control of the system parameters, including the interaction
strength \cite{Feshbach} and the particle number \cite{Hulet}, is
particularly remarkable, as it offers for the first time the
opportunity to study in greater detail fundamental issues, like
the transition from the non-interacting gas to the interacting
gas, the emergence of a condensate in going from few to many
particles, or the influence of the interaction on the structure
and the dynamics of the condensed phase.

It is easy to understand the notion of Bose-Einstein condensation
of the ideal gas. In the non-interacting system the total energy
is just the sum of energies of each particle, each particle having
at its disposal the set of single-particle energy states. Compared
with their distinguishable counterparts, indistinguishable Bose
particles have a pronounced tendency to bunch in the same
state \cite{probability}. Thus, as temperature is lowered, and the
total energy becomes limited a resource,  condensation occurs,
which is the occupation of the single-particle ground state by a
macroscopic fraction of all particles. Of course, for ultra low
temperatures also distinguishable particles would pile up in that
state. But for Bose particles, as the temperature is lowered,
condensation occurs suddenly, at a well defined temperature, and
none but only the single-particle ground state displays a truly
macroscopic population in the low temperature regime.

The picture is not so clear for interacting Bose particles.
Collisions certainly contravene the Bose particles' tendency to
pile up in a single state. In contrast to the ideal Bose gas, no
exact solution is known for the $N$-particle Hamiltonian
describing the particles in the external trapping potential (to
the delight of theorists it is a nearly perfect harmonic
oscillator potential) even if the particles' pair interaction is
modeled by a zero range pseudo-potential. Even simpler a problem,
that of a gas in a three-dimensional box with periodic boundary
conditions, is still poorly understood beyond the level of
mean-field description. Many different predictions for the
interaction induced shift of the condensation temperature, for
example, coexist in literature \cite{Tc}.

The minimal configuration where indistinguishability and
pair-interaction become manifest is a ``gas'' of two particles.
Hence, a two-particle system is the ideal starting point for a
thorough investigation of the role of indistinguishability,
the mechanism of the BEC,  and its dependence on the pair interaction
strength, without invoking the mean field theory approach.

Our discussion will be based on two distinct models. The first
model is due to Elliot H. Lieb \cite{Lieb}. It describes particles
traveling on a circumference of a circle with repulsive
$\delta$-like interaction. The exact wave functions for the $N$
atom problem are analytically known in this case, while the
eigenvalues of the energy are computed numerically from a finite
set of transcendental equations. Of course this being a
one-dimensional model, it does not display a phase transition in
the thermodynamic limit \cite{Landau}. But since we are going to
study only a case of a finite number of atoms, no non-analyticity
of any kind is expected anyway. In fact, for the purpose of
illustration we restrict our attention to the system of two atoms
only. This will prove to be sufficient to illustrate the general
strategy, which should be used in experimentally more relevant
cases. The direct extension of our calculations in the Lieb model
to more than two atoms is very tedious but is in principle
possible. The second model is that of two atoms in a
one-dimensional harmonic oscillator trap. In this case the
extension to three dimensions (3D) is simple \cite{Busch} but
going beyond 2 atoms is very difficult. In this model, the exact
eigenstates of the $N=2$ particle Hamiltonian are expressible in
terms of known special functions and the energy eigenvalues again
require a numerical solution of a transcendental equation.

Our paper is organized as follows: In
section \ref{sec:interaction} we present a general approach to the
Bose-Einstein condensation of a finite interacting system, which
is based on the properties of the single-particle density matrix.
In section \ref{sec:lieb} this general strategy is applied to the
Lieb model, while in section \ref{sec:HO} we apply this strategy
to the model of two atoms in a one-dimensional (1D) harmonic
oscillator trap. The definition of a critical temperature of our
two models is then examined in section \ref{sec:critical}.
The two final sections are devoted to concluding remarks and
acknowledgements.

\section{Interaction}
\label{sec:interaction}

Condensation in an ideal gas refers to the macroscopic occupation
of the single-particle ground state. But what is the condensate
for the interacting Bose gas? One may be tempted to call the
multi-particle ground state of the system the condensate, but this
is wrong. To repeat -- the condensate is characterized by (i) a
single-particle state which becomes (ii) macroscopically populated
in the low temperature regime. Which state could that be? Of
course, the correct definition should reduce to the one for the
ideal Bose gas when the interaction strength tends to zero. The
correct definition, it turns out, was already implicit in the
study of the coherence properties of the superfluid helium, where
the neutron scattering, probing the single-particle density matrix
has revealed the presence of an off-diagonal long-range order
\cite{Penrose}.

The single-particle density matrix derives from a generalization of the
full density matrix, $\varrho^{(1)}={\rm
Tr}'[\varrho]$, where $\varrho$ is the full density matrix of the
$N$-particle system and ${\rm Tr}'$ refers to a trace over the degrees of
freedom of a subset of $N-1$ particles. In the coordinate representation
\begin{equation}
   \varrho^{(1)}(x,x')
   =
   \int dx_2\cdots\int dx_N
   \varrho(x,x_2,\ldots,x_N;x',x_2,\ldots,x_N)\,.
\label{eq:rdm-def}
\end{equation}
As a Hermitian, positive operator with a unit trace, the single-particle
density matrix has a spectral decomposition
\begin{equation}
   \varrho^{(1)}(x,x')
   =
   \sum_{j}\lambda_{j}\phi_{j}^{*}(x)\phi_{j}(x')\,,
\label{eq:rdm_diag-def}
\end{equation}
which defines a complete set of single-particle states and corresponding
eigenvalues $\lambda_j\geq0$, $\sum_j\lambda_j =1$.

The $j$-th eigenvalue measures the weight of the corresponding
single-particle state (often called a natural orbital) in the
spectral decomposition. It may be represented as
$\lambda_{j}=\frac{\langle n_{j}\rangle}{N}$, where $\langle
n_{j}\rangle$ may be identified with the mean number of particles
in the $j$-th single-particle state. We are entitled to call one
of the states, $\phi_0$ say, {\em condensed} if the corresponding
eigenvalue is of order $O(1)$, all the other eigenvalues being
much smaller, $\lambda_j\ll\lambda_0$.

To compute the single-particle density matrix and its ensuing
spectral decomposition is not an easy task in the general case,
where the full density matrix is given in terms of the
system's statistical specifications.
For a $N$-particle system in contact
with a thermostat at temperature $T$, for example, the density
matrix reads $\varrho = \frac{1}{Z}e^{-\beta H}$, where
$\beta=1/(k_{\rm B}T)$, $H$ is the system Hamiltonian, and $Z={\rm
Tr}\varrho$ is the canonical partition function.

Being a function of the system Hamiltionian, the density matrix is diagonal
in the energy representation,
\begin{equation}
   \varrho(x_1,\ldots,x_N;x_1',\ldots,x_N')
   =
   \frac{1}{Z}\sum_{i}
   e^{-\beta E_i}\psi_i(x_1,\ldots,x_N)\psi_i(x_1',\ldots,x_N')\,,
\label{eq:cdm_erp-def}
\end{equation}
where the energy values $E_i$ and associated multi-particle wave functions
$\psi_i$ are given by the solution of the stationary Schr\"odinger equation
\begin{equation}
   H\psi(x_1,\ldots,x_N) = E\psi(x_1,\ldots,x_N)\,.
\label{eq:npart_schr-def}
\end{equation}

In the case of non-interacting particles the only correlation of
the multi-particle density matrix (\ref{eq:cdm_erp-def}) comes
from bosonic symmetrization. Hence the natural orbitals of the
single-particle reduced density matrix (\ref{eq:rdm_diag-def}) are
simply the eigenstates of the single-particle Hamiltonian.

In the case of interacting particles the situation is more complicated. Yet
the symmetry may be of help. For the system in a box with the periodic
boundary conditions, the equilibrium single-particle density matrix must be
(i) invariant under spatial translations, and (ii) it must be spatially
periodic. The spectral decomposition is therefore realized by the Fourier
expansion and the natural orbitals are just plane waves or the momentum
states,
\begin{equation}
   \rho^{(1)}(\vec{r},\vec{r}')
   =
   \sum_{\vec{p}}
   \frac{\langle n_{\vec{p}}\rangle}{N}
   \frac{e^{i\vec{p}(\vec{r}-\vec{r}')}}{L^3}\,,
\label{eq:rdm_box.res}
\end{equation}
where the momentum $\vec{p}=\frac{2\pi}{L}(j_{1},j_{2},j_{3})$
($j_{1}$,$j_{2}$, and $j_{3}$ are integer numbers). The
situation is very analogous in the standard theory of homogeneous
Bose systems which explains why the condensate is identified with
the zero momentum state.

In the inhomogeneous case, relevant for the present experiments with
magnetically trapped gases, symmetry alone cannot determine the natural
orbitals. In the most regular case of spherically symmetric trap, the natural
orbitals may be chosen proportional to the spherical harmonics (the density
matrix may be diagonalized together with the angular momentum operators), but
their radial dependence feels the interaction and in general changes with
temperature.

\section{The Lieb model}
\label{sec:lieb}

A model which to a large extent allows for an explicit
construction of the single-particle density matrix is the Lieb
model \cite{Lieb}. The Lieb model describes a collection of
interacting particles on a circle, the pair-interaction potential
being modeled by a repulsive $\delta$-function. Although this
choice of an interaction potential may seem strange it does
describe very well collisions between very slow particles, which
in a gaseous phase dominate at low temperatures, and has been
widely used in the context of BEC \cite{BEC}.

We commence with two {\em free} particles subject to the
aforementioned conditions. We introduce conveniently scaled
variables that will be used hereafter: the circle's circumference
$L$ defines the unit of length and $\hbar^{2}/2mL^{2}$ defines the
unit of energy. In scaled units the Schr\"odinger equation for two
free particles reads
\begin{equation}
   -\left[
      \frac{\partial^{2}}{\partial x_{1}^{2}} +
      \frac{\partial^{2}}{\partial x_{2}^{2}}
   \right]
   \psi(x_{1},x_{2})
   =
   E \psi(x_{1},x_{2})\,,
\label{eq:lieb_free_schreq}
\end{equation}
where $\psi(x_1,x_2)$ must obey Bose-symmetry
\begin{equation}
   \psi(x_1,x_2) = \psi(x_2,x_1)\,,
\label{eq:lieb_free_bcbose}
\end{equation}
and periodic boundary conditions,
\begin{equation}
   \psi(x_1+n,x_2+m) = \psi(x_1,x_2)\,,\qquad n,m\mbox{ integer}\,.
\label{eq:lieb_free_bcbc}
\end{equation}

\noindent Solutions of
equations (\ref{eq:lieb_free_schreq})--(\ref{eq:lieb_free_bcbc}) are
given in terms of plane waves
\begin{equation}
\label{Psi0}
   \psi(x_{1},x_{2})
   =
   \left\{
   \begin{array}{ll}
   \frac{1}{\sqrt{2}}
   \left[e^{i(k_1 x_1 + k_2 x_2)} + e^{i(k_2 x_1 + k_1 x_2)}\right]
   & \mbox{ for }k_1\neq k_2
   \\
   e^{ik_1(x_1+x_2)}
   & \mbox{ for }k_1=k_2
   \end{array}
   \right.\,,
\label{eq:lieb_free_psik1k2.res}
\end{equation}
which posses kinetic energy
\begin{equation}
   E = k_1^2 + k_2^2\,.
\label{eq:lieb_free_Ek1k2.res}
\end{equation}
Due to the boundary conditions (\ref{eq:lieb_free_bcbc}) the wave
numbers are quantized
\begin{equation}
   k_i=2\pi \nu_i\,,\qquad \nu_i=0,\pm1,\pm2,\ldots\,.
\label{eq:lieb_free_k1k2.res}
\end{equation}

\noindent As for any pair of wave numbers $(k_1,k_2)$ the wave
functions $\psi_{k_1k_2}$ and $\psi_{k_2k_1}$ refer to the very
same state, we impose the restriction
\begin{equation}
   k_1\leq k_2
\label{eq:lieb_free_k.cnd}
\end{equation}
in order to avoid the double counting of states.

We now turn to the problem of two {\em interacting} particles. The
Schr\"odinger equation pertinent to this problem reads:
\begin{equation}
   \left\{
      -\left[
         \frac{\partial^{2}}{\partial x_{1}^{2}}
         +
         \frac{\partial^{2}}{\partial x_{2}^{2}}
      \right]
      +
      2c\delta(x_{2}-x_{1})
   \right\}
   \psi(x_1,x_2)
   =
   E\psi(x_1,x_2)\,,
\label{eq:lieb_schreq}
\end{equation}
where $c$ is the strength of the interaction. Safe for the
collisions, which occur for $x_1=x_2$, we just face the
Schr\"{o}dinger equation for two {\em free} particles. Across the
line $x_1=x_2$ the first spatial derivatives of the wave function
displays a step discontinuity (because the derivative of the step
function is the $\delta$~-~function),
\begin{equation}
   \left(
      \frac{\partial}{\partial x_{2}}
      -
      \frac{\partial}{\partial x_{1}}
   \right)
   \psi\left|_{x_{2}=x_{1}^{+}}\right.
   -
   \left(
      \frac{\partial}{\partial x_{2}}
      -
      \frac{\partial}{\partial x_{1}}
   \right)
   \psi\left|_{x_{2}=x_{1}^{-}}\right.
   =
   2c\psi\left|_{x_{2}=x_{1}}\right.\,.
\label{eq:lieb_jump.cnd}
\end{equation}

\noindent It is sufficient to find a solution in the region
\begin{equation}
   S:\,
   0\leq x_{1}\leq x_{2} \leq 1\,,
\label{eq:lieb_simplex-def}
\end{equation}
which is a simplex in the $(x_1,x_2)$ plane, for if we know
$\psi(x_1,x_2)$ in $S$, we know it everywhere by symmetric,
suitable for bosons, extension.

In the simplex $S$ the problem reduces to solving the free Schr\"{o}dinger
equation (\ref{eq:lieb_free_schreq}) supplemented by the boundary conditions
\begin{equation}
   \left(
      \frac{\partial}{\partial x_{2}}
      -
      \frac{\partial}{\partial x_{1}}
   \right)
   \psi\left|_{x_{2}=x_{1}^{+}}\right.
   =
   c\psi\left|_{x_{2}=x_{1}}\right.\,,
\label{eq:lieb_Sjump.cnd}
\end{equation}
\begin{equation}
   \psi(0,x_{2})
   =
   \psi(x_{2},1)\,,\qquad
   \frac{\partial}{\partial x}
   \psi(x,x_{2})\left|_{x=0}\right.
   =
   \frac{\partial}{\partial x}
   \psi(x_{2},x)\left|_{x=1}\right.\,.
\label{eq:lieb_Sprd.cnd}
\end{equation}
Here equation (\ref{eq:lieb_Sjump.cnd}) is the adaptation of
(\ref{eq:lieb_jump.cnd}) for the simplex $S$, and
equation (\ref{eq:lieb_Sprd.cnd}) is the adaptation of the periodic
boundary conditions (\ref{eq:lieb_free_bcbc}).

One should expect the solution to the problem of two {\em
interacting} particles to be still a combination of plane waves
(as in the {\em free} case). After all, the atoms move freely most
of the time -- the only perturbation taking place when they
collide. The particular form of the solution, which indeed
respects that expectation, was proposed in a classic paper by
Elliott H. Lieb and W. Liniger in 1963 \cite{Lieb} (where the
authors dealt with the problem of $N$ interacting atoms; to date,
this so called Lieb model is the only soluble model of $N$
interacting bosons with short-range interactions). Recently,
intensively studied has been a model of interacting bosons with
the interatomic potential proportional to the product of two
particles' coordinates \cite{Gajda}. In this case the total
Hamiltonian is just a quadratic form.

In a slight deviation from the Lieb ansatz, the solution is constructed with
the help of a {\em permanent} of plane waves,
\begin{equation}
   \eta(x_{1},x_{2})
   =
   \mbox{Per}\left|\exp(ik_{i}x_{j})\right|\,.
\label{eq:eta}
\end{equation}
which is just like a determinant, yet with all minus signs turned into plus
signs. In terms of the permanent (\ref{eq:eta}), the solution reads
\begin{equation}
   \psi(x_{1},x_{2})
   \propto
   \left(
      1
      +
      \frac{c}{\partial_2 - \partial_1 -c}
   \right)
   \eta(x_{1},x_{2})\,.
\label{eq:lieb_psi}
\end{equation}
with $\partial_i\equiv\partial/\partial x_i$. As the solution appears as a
combination of plane waves, the energy eigenvalues assume the same form as in
the free case, see (\ref{eq:lieb_free_Ek1k2.res}).

The solution is specified by a pair of quasi-wave numbers $(k_1,k_2)$, the
possible values of which are determined by the boundary conditions. Inserting
the ansatz (\ref{eq:lieb_psi}) into (\ref{eq:lieb_Sjump.cnd}) and
(\ref{eq:lieb_Sprd.cnd}), these conditions turn into
\begin{eqnarray}
   K
   & = &
   \pi \mu\,,\qquad \mu=0,\pm 1, \pm 2,\ldots\,,
   \\
   k
   & = &
   2 \arctan\left(\frac{c}{2k}\right)
   + \pi \nu\,,
   \qquad \nu=0,1,2,\ldots\,.
\label{k2}
\end{eqnarray}
where $K=(k_1+k_2)/2$ and $k=(k_2-k_1)/2$. Here, the first
equation reflects the quantization of the total momentum, which is
a conserved quantity. In the second equation we refer to the
principal branch of the arcus-tangent, and we seek solutions with
$k\geq0$ to avoid double counting.

There are two interesting limiting cases: weak interaction ($c\rightarrow0$)
and strong interaction ($c\rightarrow\infty$). In the limit of weak
interaction $k\approx \sqrt{c}$ for $\nu=0$, and $k \approx
\pi \nu + c/(\nu \pi)$ for $\nu=1,2,\ldots$. These values connect smoothly
with the free case $k=\pi \nu$, $\nu=0,1,2,\ldots$. Level repulsion is
strongest for synchronized motion, i.\ e.\ when both particles travel with
approximately the same speed, $k_1\approx k_2$, and thus have ample time to ``feel'' the
collision. For the generic $N$-particle system in a box, say, the level
repulsion leads to a depletion of the density of low lying states, and the
concomitant appearance of collective excitations with a phonon-like
dispersion.

In the limit of strong interactions $k\approx(\nu+1)\pi$ for
$\nu\ll c$, and $k\approx\nu\pi$ for $\nu\gg c$. Low lying states
of the strongly interacting two-particle Bose system, for example
the ground state $\psi\propto \sin(\pi(x_2-x_1))$, can be
identified with the two-particle states of an ideal Fermi gas.
This is a particular instance of the Bose-Fermi transformation,
which maps the one-dimensional hard-sphere Bose gas onto an ideal
Fermi gas (see \cite{girar} and references therein).

In the general case the solutions for the quasi-wave number
difference $k$ may be enumerated $k_\nu$, $\nu=0,1,2,\ldots$, the
$\nu$-th solution being strictly bounded $\pi\nu \leq
k_\nu<\pi(\nu+1)$. Expressed in terms of sums and differences of
quasi-wave numbers, the spectrum of energy eigenvalues is given by
(see equation (\ref{eq:lieb_free_Ek1k2.res}))
\begin{equation}
   E_{\mu\nu}
   =
   2k_\nu^{2}
   +
   2\pi^2\mu^2\,,
   \qquad \nu=0,1,\ldots\,,
   \qquad \mu=0,\pm1,\ldots\,.
\label{eq:lieb_Emunu.res}
\end{equation}
The corresponding eigenfunction may be also labeled,
$\psi_{\mu\nu}(x_1,x_2)$, where integer $\mu$ specifies the
center-of-mass motion, and non-negative integer $\nu$ specifies
the relative motion.

We are now in a position to study the system statistical
properties. Adapting the defining equation (\ref{eq:cdm_erp-def})
to our present situation, and considering for the time being only
states with a zero total momentum, $\mu=0$, the system canonical
density matrix is given by
\begin{equation}
   \varrho(x_1,x_2;x_1',x_2')
   =
   \sum_{\nu=0}^{\infty} p_{\nu}\varrho_{\nu}(x_1,x_2;x_1',x_2')\,,
\label{eq:lieb_cdm-def}
\end{equation}
where $p_{\nu}$ is the Boltzmann weight,
\begin{equation}
   p_{\nu} = \frac{1}{Z}e^{-\beta E_{0\nu}}\,,
\label{eq:lieb_p0nu-def}
\end{equation}
and $\varrho_{\nu}$ is the pure state density matrix,
\begin{displaymath}
   \varrho_{\nu} = \psi_{0\nu}(x_1,x_2)\psi_{0\nu}^{*}(x_1',x_2')\,.
\end{displaymath}

It is instructive to compute the reduced single-particle density
matrix $\varrho^{(1)}_{\mu\nu}$, see equation (\ref{eq:rdm-def}) for
the definition. After some tedious yet straightforward calculation
one obtains
\begin{eqnarray}
   \varrho^{(1)}_{\mu\nu}(x,x')
   &=&
   \frac{(-)^\mu\sin k_\nu
      - k_\nu[1-(-)^\mu\cos k_\nu]z}
   {k_\nu+(-)^\mu\sin k_\nu} \cos(k_\nu z) \nonumber \\
   &+&
   \frac{[1-(-)^\mu\cos k_\nu] + z(-)^\mu k_\nu\sin k_\nu}
   {k_\nu+(-)^\mu\sin k_\nu} \sin(k_\nu z)
\label{eq:lieb_rdm_nu.res}
\end{eqnarray}
with $z=|x-x'|$.
The thermal equilibrium reduced density matrix is given by
\begin{displaymath}
   \varrho^{(1)}(x,x')
   =
   \sum_{\nu=0}^{\infty}
   p_{\nu}\varrho^{(1)}_{\nu}(x,x')\,.
\end{displaymath}
According to the discussion in section \ref{sec:interaction}
it may be diagonalized
\begin{displaymath}
   \varrho^{(1)}(x,x')
   =
   \sum_{j=-\infty}^{\infty}
   \lambda_j
   \phi_j(x)\phi_j^{*}(x'),
\end{displaymath}
where $\lambda_j=\frac{\langle n_j\rangle}{2}$, with $\langle
n_j\rangle$ the mean occupation of the single-particle state
$\phi_j(x)$. Due to the translational invariance of the Lieb
model, the single-particle states are plane waves, $\phi_j(x) =
e^{2\pi i j x}$, $j=0,\pm1,\pm2,\ldots$.

The state with the largest eigenvalue in this decomposition will be referred
to as the single-particle ground state and denoted by $\phi_{0}$. In the
present model $\phi_0(x)=1$ -- most simple a wave function indeed.

We shall be interested in the mean and root-mean square
fluctuations of the single-particle ground-state occupation,
\begin{equation}
   \xi
   =
   \sqrt{\langle n_{0}^{2}\rangle-\langle n_{0}\rangle^{2}}\,.
\label{fluct0}
\end{equation}
The mean and mean-square occupation of the single-particle ground state can
be expressed as
\begin{equation}
   \langle n_{0}\rangle
   =
   2\cdot P_{2} + 1\cdot P_{1} + 0\cdot P_{0}\,,
\label{eq:lieb_mean0}
\end{equation}
\begin{equation}
   \langle n_{0}^{2}\rangle
   =
   4\cdot P_{2} + 1\cdot P_{1} + 0\cdot P_{0}\,,
\label{eq:lieb_var0}
\end{equation}
where $P_{n}$ denotes probability to find $n$ atoms in the single-particle
ground state.

The probability to find two particles in the single-particle ground state
can of course not be calculated from the single-particle density matrix, but
must rather be inferred from the full density matrix,
$P_2=\langle\phi_0\phi_0|\varrho|\phi_0\phi_0\rangle$. In the coordinate
representation,
\begin{displaymath}
   P_2
   =
   \sum_{\nu}
   p_{0\nu}\left|
      \int dx_1 dx_2
      \phi_0^{*}(x_1)\phi_0^{*}(x_2)
      \psi_{0\nu}(x_1,x_2)
   \right|^{2}\,.
\end{displaymath}
In order to compute $P_1$ we recall $\langle n_{0}\rangle = 2
\langle\phi_0|\varrho^{(1)}|\phi_0\rangle$ by definition, which in
combination with (\ref{eq:lieb_mean0}) yields
\begin{displaymath}
   P_1
   =
   2\int dx dx'
   \phi_0^{*}(x)\varrho^{(1)}(x,x')\phi_0(x')
   -
   2 P_2\,.
\end{displaymath}
Finally, as the statistical events to find zero, one or two
particles in the single-particle ground state are mutually
exclusive and complete,
\begin{displaymath}
   P_0 = 1 - P_1 - P_2\,.
\end{displaymath}

We note that in a restricted ensemble with zero total momentum the
probability to find exactly one particle in the single-particle
ground state is strictly zero, $P_1=0$. For if one particle would
be at rest, the other particle must be at rest too. Hence, in such
a restricted ensemble, only $P_2$ and $P_0$ can be different from
zero.

Before we continue with the description of our results, we must
supply a few words about technicalities of this work. All of them
regard computation of the thermal density matrix
(\ref{eq:lieb_cdm-def}). In order to compute the partition
function $Z$ we sum over the first $100$ states. The number of
states used in the construction of the thermal density matrix is
chosen so that the sum of all probabilities $p_{\nu}$
(\ref{eq:lieb_p0nu-def}) differs from $1$ by less than $0.01$.
This usually renders a number between $\nu_{max}=1$ (for very low
temperatures) and $\nu_{max}=6$ (for relatively high temperatures,
$T\approx 1000$). This choice constitutes an important aspect as
the time of computation grows very quickly with the number of
states involved.

\begin{figure}[htbp]
\begin{center}
\includegraphics[width=2.5in]{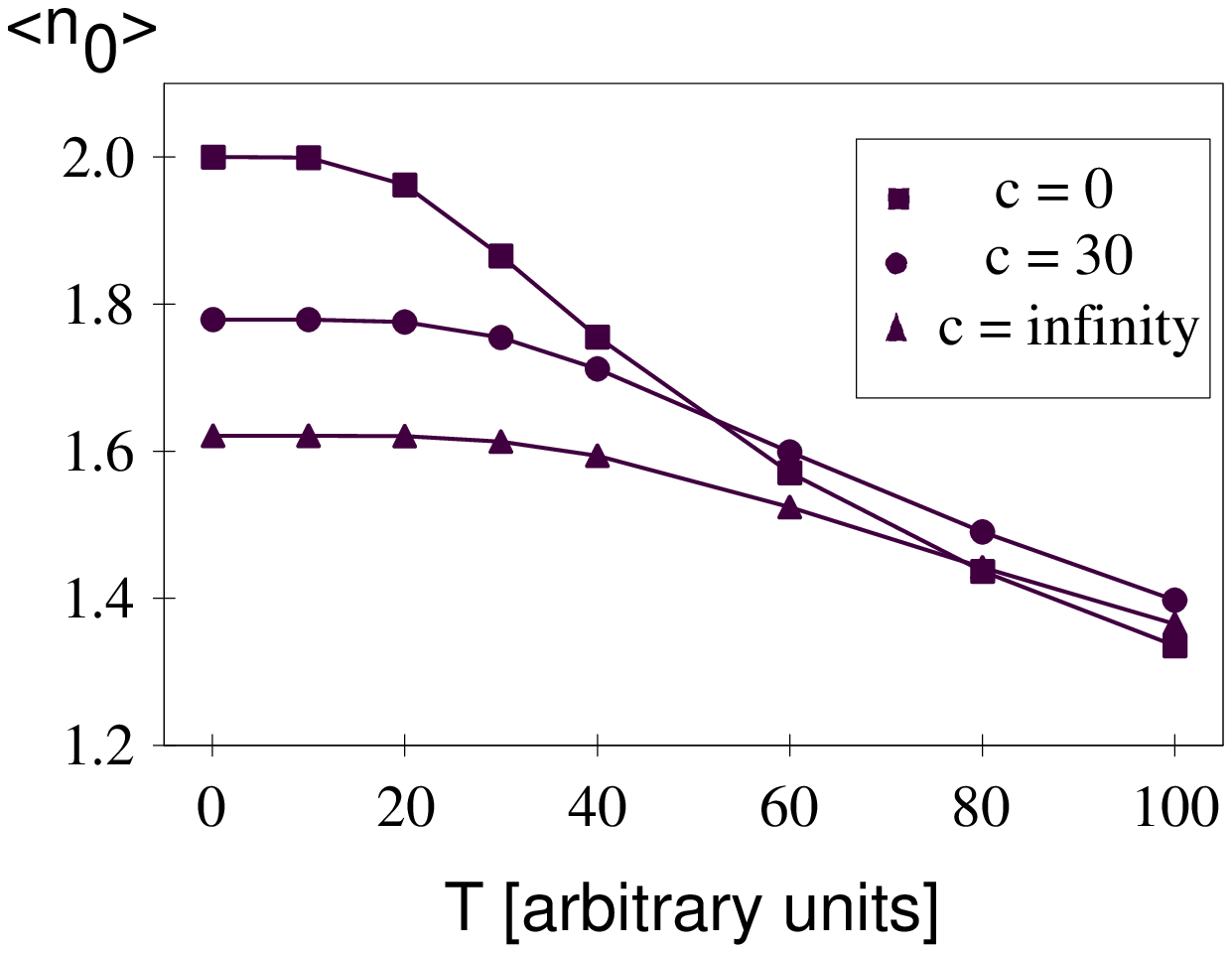}
\caption{Temperature dependence of the mean ground-state
occupation in the Lieb model for the free case($c=0$, squares) as
well intermediate ($c=30$, circles) and infinite interaction
($c\rightarrow \infty$, triangles).\label{fig1}}
\end{center}
\end{figure}

\begin{figure}
\begin{center}
\includegraphics[width=2.5in]{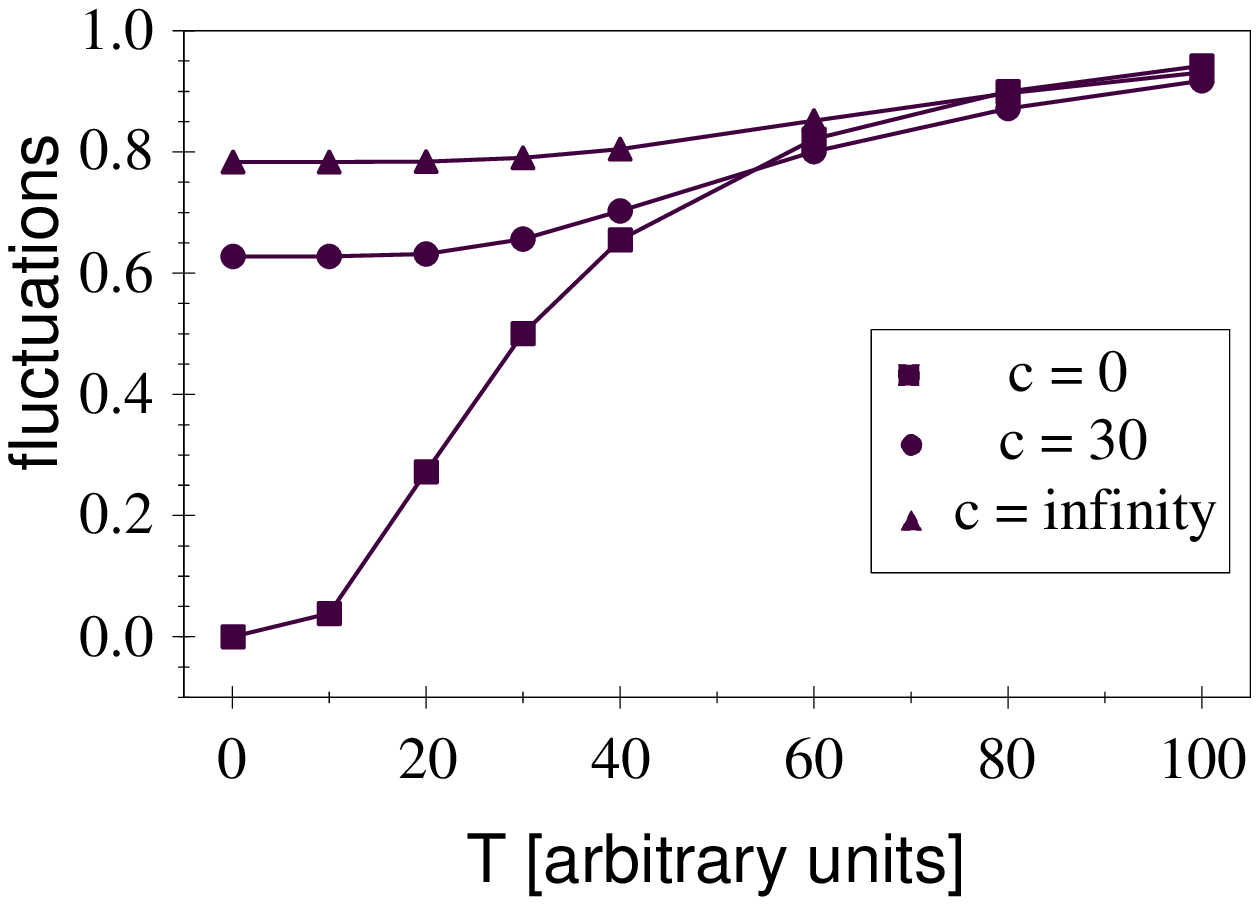}
\caption{Temperature dependence of the fluctuations of the
ground-state occupation in the Lieb model for the free case
($c=0$, squares) as well intermediate ($c=30$, circles) and
infinite interaction ($c\rightarrow \infty$, triangles).
\label{fig2}}
\end{center}
\end{figure}

Now we can analyze the dependence of the occupations, fluctuations
and probabilities  on temperature and interaction strength
(defined by the magnitude of $c$). Figure \ref{fig1} presents the
dependence of mean occupation of the ground state (\ref{eq:lieb_mean0})
on temperature for various values of the interaction magnitude
$c$. Obviously, in the free case ($c=0$) at $T=0$ both atoms are
in the 1-particle ground state (which can be traced back to its
definition). For $c \neq 0$ mean occupation of the ground state at
$T=0$ is still substantial, but certainly smaller than $2$ and
assumes the smallest possible value for $c\rightarrow \infty$.
Interestingly, for higher temperatures ($T\approx 50$) $\langle
n_{0}\rangle$ can be greater in interacting cases than in the free
one. Analogous behavior can be observed in the dependence of
fluctuations (\ref{fluct0}) on temperature. Here only for the free
case they can take on the zero value, but again they can be bigger
than in interacting cases for higher temperatures (see figure
\ref{fig2}).

\section{Two atoms in a harmonic trap}
\label{sec:HO}

Before we turn our attention to an example of two interacting
bosons in a harmonic trap, let us comment briefly on the
non-interacting case. The Hamiltonian describing two
non-interacting particles in a one-dimensional harmonic trap takes
a very simple form:

\begin{displaymath}
H=\hbar \omega (a_1^{\dagger}a_1+a_2^{\dagger}a_2)\: ,
\end{displaymath}
where $a_1$ and $a_2$ are lowering operators of the first and the
second particle, respectively. The canonical-ensemble density
matrix is

\begin{displaymath}
\rho ={1 \over Z}\exp \left[ -{H \over {kT}}\right] \: ,
\end{displaymath}
where the partition function $Z$ ensures the normalization of the
density matrix. Until now we have not specified whether we deal
with distinguishable or indistinguishable particles. This last
property determines the space of available states. For
distinguishable particles the trace defining the partition
function extends over all configurations of particles
independently occupying the levels:

\begin{displaymath}
Z_{cl}=\sum\limits_{n_1,n_2} {q^{n_1+n_2}=(1-q)^{-2}} \: ,
\end{displaymath}
where $q=\exp [-{{\hbar \omega }/{kT}}]$. For identical bosons
there are fewer states since the quantum identical particles
cannot be labeled. So the sum in the trace is restricted:

\begin{displaymath}
Z_B=\sum\limits_{n_2=0}^\infty
{q^{n_2}}\sum\limits_{n_1=n_2}^\infty {q^{n_1}}=[(1-q)(1-q^2)]^{-1}.
\end{displaymath}

In both models discussed here we compute the probability of
finding zero, one, and two particles in the "condensed" state.
This is easily done for classical and quantum particles in the
present example. The only tool needed is the sum of the geometric
series. One easily gets

\begin{eqnarray*}
P^{cl}_{0}&=&q^2  \\ P^{cl}_1&=&2q(1-q) \\ P^{cl}_2&=&(1-q)^2
\end{eqnarray*}
for distinguishable particles and

\begin{eqnarray*}
P^B_0&=&q^2 \\ P^B_1&=&q(1-q^2) \\ P^B_2&=&(1-q)(1-q^2)
\end{eqnarray*}
for bosons. Knowing the probability distributions we may easily
compute all expectation values. The simplest is just the
temperature dependence of mean number of particles in the ground
state. In the two cases considered we get

\begin{eqnarray*}
\langle n_0 \rangle^{cl}&=&2(1-q) \\ \langle n_0
\rangle^B&=&(2-q)(1-q^2)
\end{eqnarray*}
One can easily check that for all finite temperatures $\langle
n_0\rangle^{B}>\langle n_0\rangle^{cl}$. In other words, as the
temperature decreases, the population of the ground state grows
faster for bosons due to the restricted space of states. Our
reader should have no problem extending the results presented so
far to the case of N non-interacting particles in a 1D harmonic
trap. For N going to infinity this is the origin of the
Bose-Einstein phase transition.

We investigate now the behaviour of two identical, interacting
bosonic atoms in a harmonic trap. Again we approximate atom-atom
interactions with a repulsive potential of a zero range. This
system has analytic solutions which have been derived in
\cite{Busch}, where the three-dimensional case is discussed in
greater detail. We summarize here the main results for the
one-dimensional system.

The Hamiltonian can be decomposed into two parts:

\begin{displaymath}
H=H_{cm}+H_{rel},
\end{displaymath}
where $H_{cm}$ is the Hamiltonian of the center of mass, whose
coordinate is defined here as $X=\sqrt{1/2}(x_{1}+x_{2})$, and
$H_{rel}$ is the Hamiltonian of the relative coordinate, defined
here as $x=\sqrt{1/2}(x_{1}-x_{2})$. This unconventional choice of
the two coordinates differs from the usual definitions by a
numeric factor and makes the effective masses $M$ of the two
degrees of freedom equal. The Hamiltonian of the center of mass
$X$ is equal to the Hamiltonian of the harmonic oscillator,

\begin{displaymath}
H_{cm}=-\frac{\hbar^2}{2M}\nabla^2_{X}+\frac{1}{2}M \omega^2 X^2,
\end{displaymath}
whereas the Hamiltonian for the relative coordinate is

\begin{displaymath}
H=-\frac{\hbar^2}{2M}\nabla^2_{x}
+\frac{1}{2}M\omega^2x^2+2\widetilde{c}\delta(x) \, .
\end{displaymath}
Here the last term, proportional to Dirac's $\delta$-function,
describes the zero-range interaction. The parameter
$\widetilde{c}$ determines the intensity of the zero-range
interaction. For $\widetilde{c}=0$, we have two free atoms. To
introduce dimensionless coordinates we scale both $x$ and $X$ by
the oscillator length $d=\sqrt{\frac{M\omega}{\hbar}}$. Now the
Hamiltonian of the relative coordinate takes the following form:

\begin{displaymath}
H_{rel}=\hbar\omega(-\frac{1}{2}\nabla^2+\frac{1}{2}x^2 +2c\delta
(x)) \, ,
\end{displaymath}
where the interaction strength depends on the dimensionless
parameter $c$. Since we consider repulsive interactions, $c$ must
be greater than $0$.

The eigenvalues and eigenfunctions of the center of mass are well
known:

\begin{displaymath}
E_{n,cm}=\hbar\omega(n+\frac{1}{2})
\end{displaymath}
with $n=0,1,2,\ldots$, and

\begin{displaymath}
\psi_{n}\left( X \right)= \frac{1}{\pi^{1/4}}\frac{1}{\sqrt{2^n
n!}} e^{-\frac{X^2}{2}} H_{n}\left( X \right).
\end{displaymath}

\noindent All these eigenfunctions are symmetric under the
exchange of the two particles.

For the relative coordinate we have two different sets of
solutions. Half of them are odd eigenfunctions which vanish for
$x=0$. Therefore they are not influenced by the zero-range
interaction and reduce to the usual solutions of the harmonic
oscillator. We do not consider them, because they are odd
functions of the relative coordinate and therefore their
combination with any eigenstate of the center of mass describes
fermions instead of bosons (incidentally, we note that
two trapped fermions do
not feel the zero-range interaction, in accordance with the well
known fact that cold fermions are insensitive to $s$-wave
scattering). The
remaining eigenfunctions are:

\begin{displaymath}
\psi_{\nu}\left( x \right)= \frac{B}{\sqrt{\pi}}\Gamma(-\nu)
e^{-\frac{x^2}{2}} U\left( -\nu, \frac{1}{2}, x^2\right) \, ,
\end{displaymath}
where

\begin{displaymath}
B^2=\frac{\Gamma(\frac{1}{2}-\nu)}
{\Gamma(-\nu)[\psi(\frac{1}{2}-\nu)-\psi(-\nu)]}
\end{displaymath}
(here $\psi$ is the logarithmic derivative of the gamma function
$\Gamma$) and $U$ is the confluent hypergeometric function (see page
504 of reference \cite{abrste}). These eigenfunctions are symmetric
under the exchange of the two atoms, so we can use them in
conjunction with the eigenstates of the center of mass to describe
two bosonic atoms. The values of energy are:

\begin{displaymath}
E_{n,rel}=\hbar\omega(\nu_{n}+\frac{1}{2}) \;\;\; n=0,2,4,\ldots
\, ,
\end{displaymath}
where the non-integer parameters $\nu$ are solutions of the
transcendental equation

\begin{displaymath}
c\Gamma(-\nu)=-\Gamma(1/2-\nu) \, .
\end{displaymath}
They also satisfy the condition $n<\nu_{n}<n+1$ and

\begin{eqnarray}
\lim_{c\rightarrow 0}\nu_{n} & = & n \;\;\; \mbox{(no
interaction)} \, , \\ \lim_{c\rightarrow \infty}\nu_{n} & = & n+1
\;\;\; \mbox{(infinite interaction)}.
\end{eqnarray}
The eigenstates of the system are therefore:

\begin{displaymath}
\Psi_{n,\nu} (x_{1},x_{2}) = \psi_{n} (X) \psi_{\nu} (x) \, .
\end{displaymath}

\begin{figure}
\begin{center}
\includegraphics[width=2.5in,clip]{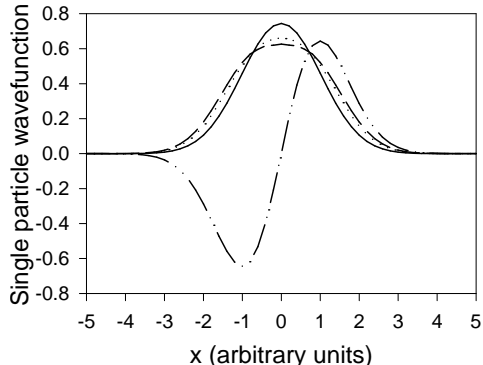}
\caption{Single-particle ground-state wave
functions of 2 atoms in a 1D harmonic trap at $T=0$ for $c=0$
(solid line), $c=1$ (dotted line), $c=2$ (dashed line), and
$c=\infty$ (dash-dotted line).
\label{fig3}}
\end{center}
\end{figure}

We can now build the density matrix for the two trapped atoms. At
$T=0$ the density matrix is simply given by the ground state of
the system

\begin{displaymath}
\rho(x_{1},x_{2};x_1',x_2')= \Psi_{0,\nu_{0}}^{*}(x_{1},x_{2})
\Psi_{0,\nu_{0}}(x_1',x_2')
\end{displaymath}
and the one-particle density matrix is

\begin{equation} \label{diagonalization}
\rho^{(1)}(x,x')=\int dy \rho(x,y;x',y)=
\sum_{i=0}^{\infty}\lambda_{i}\phi_{i}(x) \phi_{i}(x')
\end{equation}
The integration and the diagonalization as indicated in
(\ref{diagonalization}) must be done numerically. The {\em
single-particle ground-state wave function} $\phi_{0}$ is plotted
in figure \ref{fig3} for different values of $c$. These states
differ from the ground state of the harmonic oscillator more and
more as the interaction increases.

For $T>0$ we work again in the framework of the canonical
ensemble. The density matrix is

\begin{eqnarray}
\rho(x_{1},x_{2};x_1',x_2') & = &
\frac{\exp[-\frac{x_{1}^{2}+x_{2}^{2}+x_1'^{2}+x_2'^{2}}{2}]}
{\pi^{3/2}} \sum_{n,\nu}
\frac{e^{-\frac{\hbar\omega(n+\nu)}{k_{B}T}}}{Z}
\frac{\Gamma(\frac{1}{2}-\nu)\Gamma(-\nu)}
{2^{n}n![\psi(\frac{1}{2}-\nu)-\psi(-\nu)]} \nonumber \\ & & \times
H_{n}\left( \frac{x_{1}+x_{2}}{\sqrt{2}}\right) H_{n}\left(
\frac{x_1'+x_2'}{\sqrt{2}}\right) \nonumber \\ & & \times
U\left( -\nu, \frac{1}{2}, \frac{(x_{1}-x_{2})^2}{2}\right)
U\left( -\nu, \frac{1}{2}, \frac{(x_1'-x_2')^2}{2}\right) \, .
\end{eqnarray}
In this case, technical difficulties come from the numerical
evaluation of the $U$-functions with large values of $\nu$,
especially for moderate and strong interactions. This problem
restricts our investigation to low temperatures (between 0 and 6
nanokelvin, for a typical trap frequency $\omega=2\pi\cdot 100$
Hz). The partition function is given by the sum over the first 90
states, and again (like for the Lieb model) the density matrix is
constructed so that the sum of probabilities
$p_{n\nu}=\exp(-\frac{\hbar\omega(n+\nu)}{k_{B}T})/Z$ differs from
1 by less than $0.01$. The prescription for the calculation of
occupations, fluctuations and probabilities $P_{0}$, $P_{1}$,
$P_{2}$ is the same as in the previous section.

Note that, unlike in the translationally invariant case of the
Lieb model, the natural orbitals (eigenvectors of the
single-particle density matrix) depend here both on the
temperature and on the interaction strength. The temperature
dependence turns out to be relatively weak. In particular, the
ground state $\phi_{0}$ of the thermal reduced density matrix is
very similar to that at $T=0$. This is consistent with the fact
that the Gross-Pitaevskii equation \cite{GP} (used to determine the
wave function of the weakly-interacting condensate within the so
called mean-field approximation), which is obtained for $T=0$, is
a very good tool also for the conditions realized in the
experiments, where $T>0$ inevitably.

\begin{figure}
\begin{center}
\includegraphics[width=2.5in,clip]{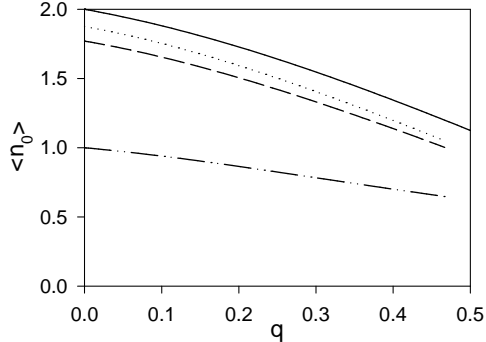}
\caption{Mean occupation of the single-particle
ground state of 2 atoms in a 1D harmonic trap vs.
$q=\exp(-\hbar\omega /kT)$ for $c=0$ (solid line), $c=1$ (dotted
line), $c=2$ (dashed line), and $c=\infty$ (dash-dotted line).
\label{fig4}}
\end{center}
\end{figure}

The mean occupation of the single-particle ground state is plotted
in figure \ref{fig4}. In analogy to the Lieb model, $\langle n_{0}
\rangle$ at $T=0$ reduces from 2 (no interaction) to lower values
as $c$ increases. However, the curves for different values of $c$
do not cross each other.

\begin{figure}
\begin{center}
\includegraphics[width=2.5in,clip]{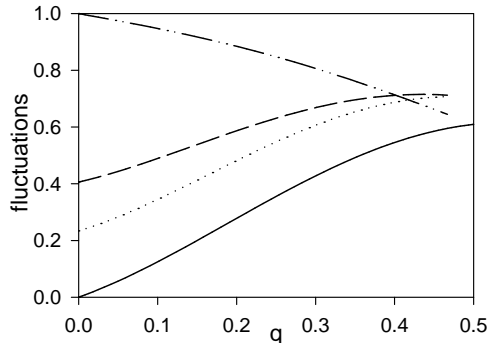}
\caption{Fluctuations of the ground-state   
occupation for 2 atoms in a 1D harmonic trap vs.
$q=\exp(-\hbar\omega /kT)$ for $c=0$ (solid line), $c=1$ (dotted
line), $c=2$ (dashed line), and $c=\infty$ (dash-dotted line).  
\label{fig5}}
\end{center} 
\end{figure}

The fluctuations of the ground-state occupation are depicted in
figure \ref{fig5}. The curves for different values of $c$ seem to
cross beyond the range of temperatures we have investigated. It is
remarkable that for infinite interaction the fluctuations are a
decreasing function of temperature, contrarily to the case of low
$c$.

\section{The critical temperature}
\label{sec:critical}

The critical temperature $T_{c}$ defines a range of temperatures
$0<T\le T_{c}$ where a macroscopic fraction of the gas occupies the
single-particle ground state. It is therefore reasonable to define
the critical temperature in our two-atom systems by means of the
crossing between $P_{0}$ and $P_{1}$ or $P_{2}$, in the same spirit
of the approach presented in \cite{faber} for a weakly interacting
Bose gas.

\subsection{$T_{C}$ for the Lieb model}

\begin{figure}
\begin{center}
\includegraphics[width=2.5in]{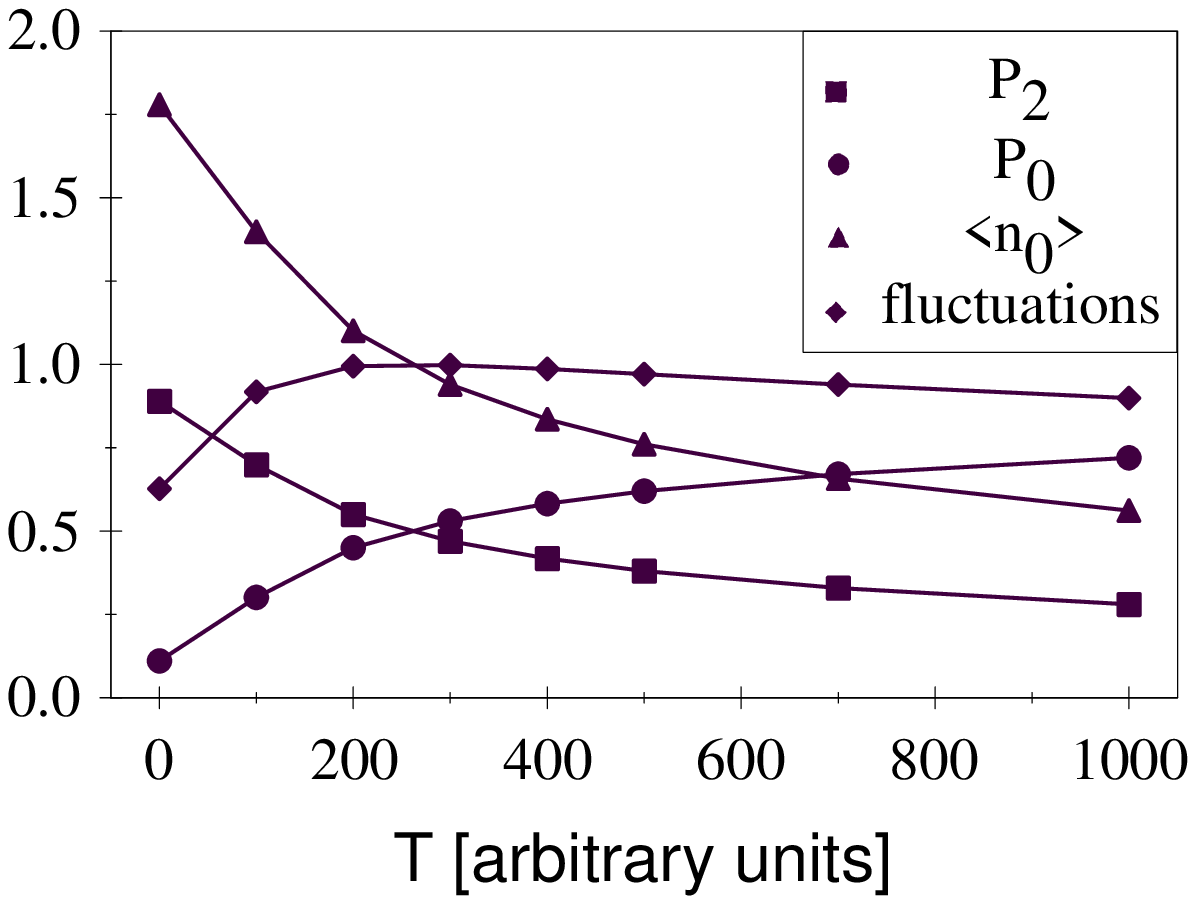}
\caption{Temperature dependence of probabilities $P_{2}$ and
$P_{0}$ as well as of the mean and fluctuations of the
ground-state occupation for the case $c=30$ in the Lieb model.
\label{fig6}}
\end{center}
\end{figure}

As regards the probabilities of populating the ground state by a
certain number of atoms ($P_{0}$,$P_{1}$ and $P_{2}$), our first
observation is that for any interaction strength there is a
crossing between $P_{2}$ and $P_{0}$ for a non-zero temperature
(see figure \ref{fig6} -- the sample case of $c=30)$. It defines
some characteristic temperature $T_{c}$ below which the ground
state is populated by 2 rather than no particles. We believe it is
an analogue of the critical temperature for higher-dimensional
systems in the thermodynamic limit. This crossing corresponds to
the maximum of fluctuations ($\xi=1$) as well as to the point
where the mean occupation of the ground state $\langle
n_{0}\rangle$ crosses this maximum becoming smaller than
fluctuations for temperatures higher than $T_{c}$. All these
features are illustrated in figure \ref{fig6}.

\begin{figure}
\begin{center}
\includegraphics[width=2.5in]{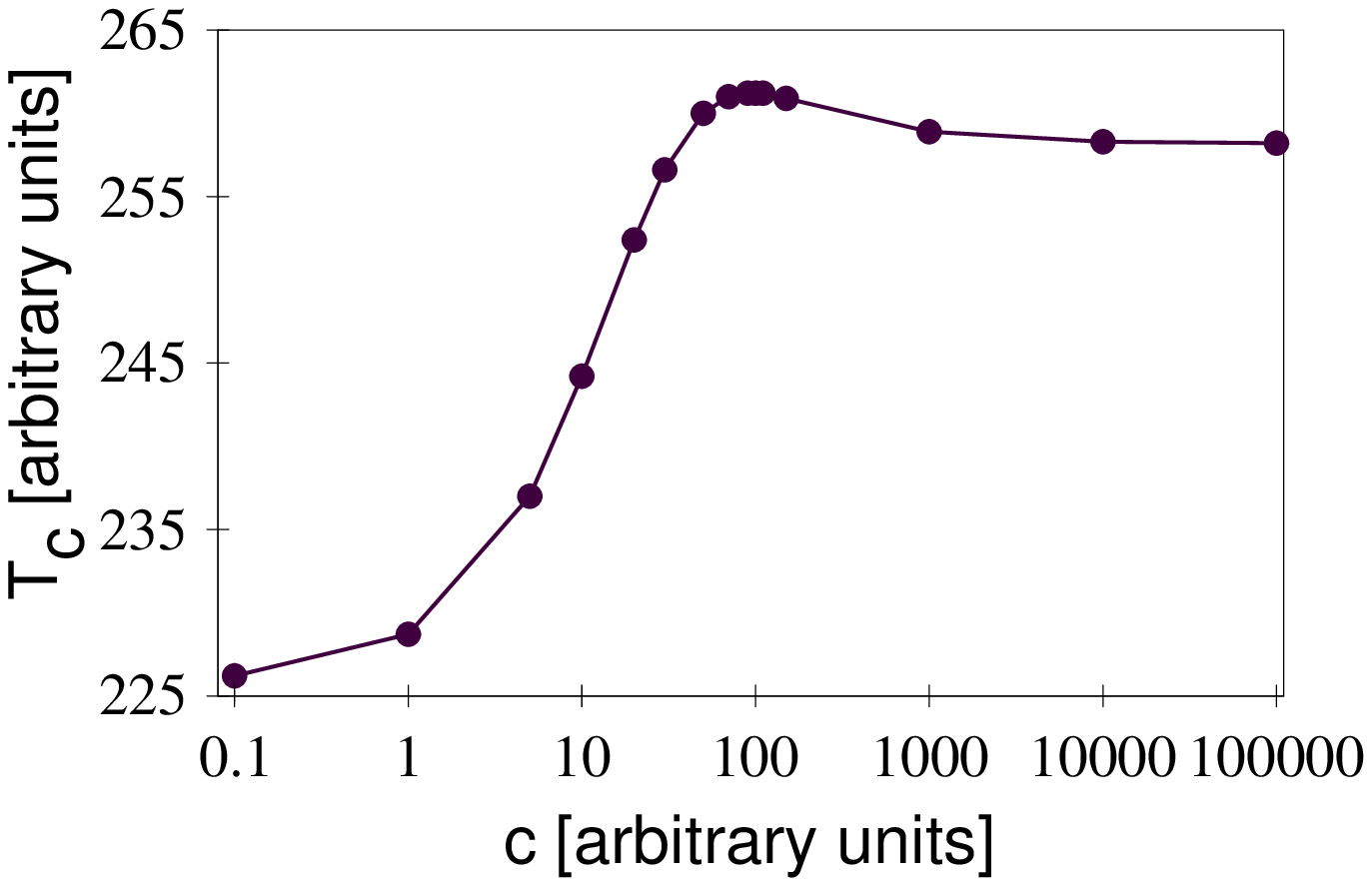}
\caption{Dependence of the characteristic temperature $T_{c}$ on
the interaction strength $c$ -- case of zero total momentum in the
Lieb model.
\label{fig7}}
\end{center}
\end{figure}

The most interesting aspect of the dependence of the
characteristic temperature $T_{c}$ on the interaction strength is
whether it increases or decreases as interaction magnitude grows.
This relation is presented in figure \ref{fig7} for all positive
$c$ values. We observe that initially the characteristic
temperature increases with the interaction strength, but having
reached a maximum value for $c\approx 100$ it decreases thus
rendering some characteristic interaction magnitude.
Such a maximum occurs also in real condensates
\cite{rep}.

\begin{figure}
\begin{center}
\includegraphics[width=2.5in]{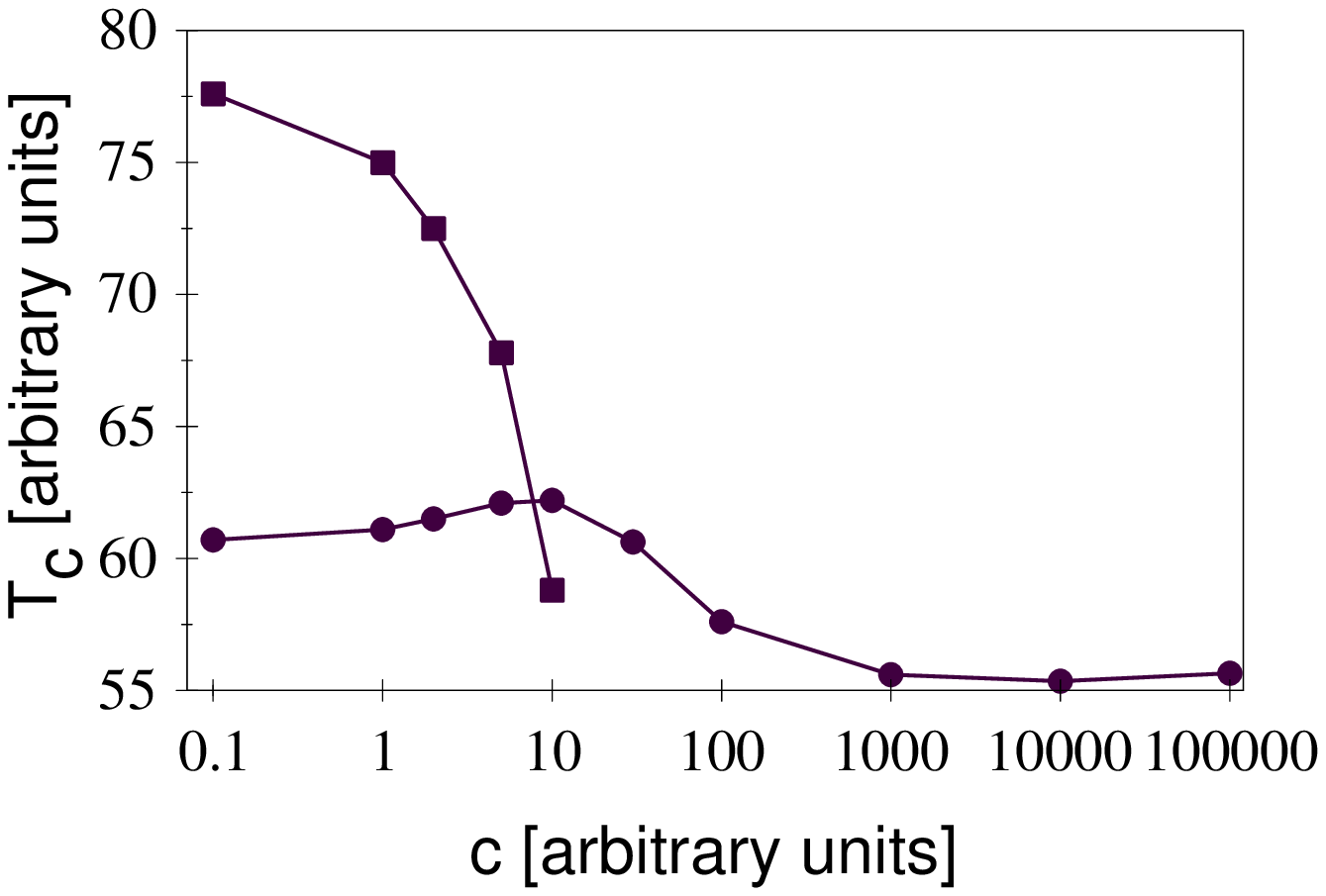}
\caption{Dependence of the characteristic temperature $T_{c}$ on
the interaction strength $c$ -- case of any total momentum in the
Lieb model. Circles indicate crossings between $P_{2}$ and $P_{0}$
whereas squares -- between $P_{1}$ and $P_{0}$.
\label{fig8}}
\end{center}
\end{figure}

We should remind ourselves the special condition of a zero total
momentum imposed on the system. It is automatically fulfilled in
the canonical ensemble for $N \rightarrow \infty$. However, as
here $N=2$ we should also look into the case when the total
momentum is unconstrained. There are several important new
features in this case -- $P_{1}$ is no longer zero and it may also
cross $P_{0}$ and $P_{2}$. It is tempting to investigate {\em two}
characteristic temperatures now: one defined by the crossing of
$P_{2}$ with $P_{0}$ and the other one when $P_{1}$ crosses
$P_{0}$. Their dependence on the interaction magnitude $c$ is
presented in figure \ref{fig8}. Circles indicate crossings between
$P_{2}$ and $P_{0}$ whereas squares -- between $P_{1}$ and
$P_{0}$. Now, as we see, for small $c$ $P_{1}$ - $P_{0}$ is met
first when going from high to lower temperatures. At some $c$,
$P_{2}$ - $P_{0}$ temperature is higher. From some $c$ between
$10$ and $20$ up there is no $P_{1}$ - $P_{0}$ crossing -- simply
$P_{0}$ is always greater than $P_{1}$. If we follow the higher of
the two temperatures as $c$ grows, we notice that this quantity
almost always decreases (except for very large $c$ values). This
result is opposite to the case of zero total momentum.

\subsection{$T_{C}$ for the trapped atoms}

\begin{figure}
\begin{center}
\includegraphics[width=2.5in,clip]{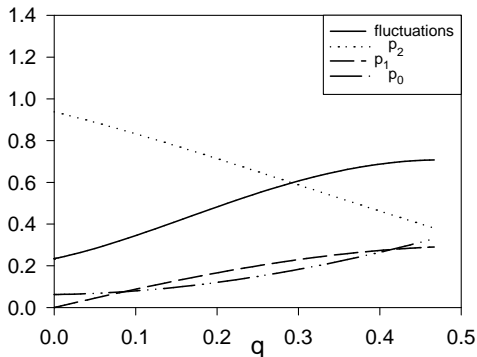}
\caption{Fluctuations of the ground-state
occupation (solid line) and probabilities $P_{2}$, $P_{1}$ and
$P_{0}$ for 2 atoms in a 1D harmonic trap (dotted, dashed, and
dash-dotted lines, respectively) vs. $q=\exp(-\hbar\omega /kT)$
for $c=1$.
\label{fig9}}
\end{center}
\end{figure}

 The correspondence between the maximum of fluctuations and
the crossing between $P_{2}$ and $P_{0}$ that we have noted in the
Lieb model might exist also for two bosons in a harmonic trap.
Unfortunately it occurs at temperatures that are beyond the range
of our investigation. Only for $c=\infty$ and $c=0$ can this
correspondence be safely confirmed. However, a qualitative
extrapolation of our data (see figure \ref{fig9} as an example)
supports this conclusion.

\begin{figure}
\begin{center}
\includegraphics[width=2.5in,clip]{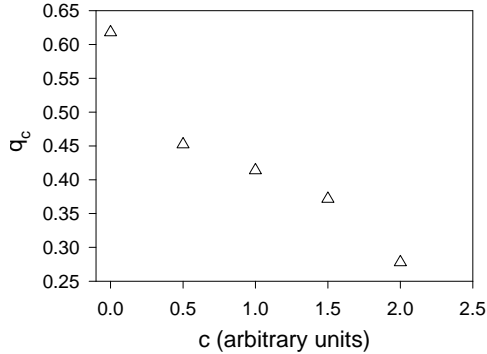}
\caption{Critical values $q_{c}=\exp[-\hbar
\omega/kT_{c}]$ vs. $c$ for 2 atoms in a 1D harmonic trap.
\label{fig10}}
\end{center}
\end{figure}

For two bosons in a trap the probability $P_{1}$ does not vanish
at $T\neq 0$, so the probabilities $P_{1}$ and $P_{0}$ cross at
some characteristic temperature $T_{c1}$, although only for
moderate interactions ($c\le 2$). This critical temperature
decreases from about 12 nK ($c=0$) to about 3.8 nK ($c=2$) for the
used parameters. The crossing between $P_{2}$ and $P_{0}$ occurs
for any value of $c$ and the critical temperature $T_{c2}$ goes
from about 8.1 nK for free bosons to $T=0$ for $c\rightarrow
\infty$ (see figure \ref{fig10}). We note the different behaviour
of critical temperature when compared to the Lieb model.

\section{Conclusions}

In this paper we have presented a detailed analysis of the
statistical properties of two interacting bosons in one dimension.
The principal advantage of our work lies in the fact that the
general strategy could be carried out exactly in this case leaving
no doubts on the nature of the approximations that would have to
be employed otherwise. Still, the calculations involved do not
present substantial mathematical difficulties permitting a
complete review of the methods and results.

The situation in a real experiment is much more complex. In recent
experiments $2000$ to $5 \cdot 10^6$ atoms have been condensed in
a nearly perfectly harmonic trap. While the statistical properties
of the finite ideal Bose gas are well understood within the
framework of the restricted ensembles (canonical \cite{canonical}
and microcanonical \cite{microcanonical}), the impact of
interactions is far from clear. Most phenomena are sufficiently
well described within the mean-field approximation, suitable for
the condensate at zero temperature. Finite temperature phenomena
are significantly harder to compute. We have already mentioned
conflicting predictions of the shift of the critical temperature
\cite{Tc}. Another example is -- as yet -- unexplained temperature
dependence of the measured eigenfrequencies of the condensate
\cite{excitations}.

We believe that a clear presentation of the fundamental concepts,
applied to soluble models may shed some light on the problem of
interacting cold bosons at a real-world scale.

\section{Acknowledgments}

M A C gratefully acknowledges financial support from MURST 60\%
'atoms and radiation' and thanks the Center for Theoretical
Physics for friendly hospitality. K Rz and K G  acknowledge the
support of the subsidy from the Foundation for Polish Science and
of the Polish KBN Grant no 2P03B05715. M A C  was also partially
supported by the Polish KBN Grant no 2P03B13015.
M W gratefully acknowledges financial support  
from DFG Programme SPP 1116
``Wechselwirkung in ultrakalten Atom- und  
Molek\"ulgasen''.


\begin{thebibliography}{99}
\bibitem{probability}
Consider, for example, $2$ particles which are statistically
distributed over $M$ states. For distinguishable particles there
are in total $C_{\rm tot}=M^2$ configurations with
double-occupancy occurring $C_{\rm dble}=M$ times. For
indistinguishable particles, on the other hand, there are in total
$C_{\rm tot}=M(M+1)/2$ configurations, $M(M-1)/2$ single occupancy
configurations, and $C_{\rm dble}=M$ double occupancy
configurations. The a priori probability to find a
double-occupancy is given by $P_{\rm dble}^{\rm dst}=1/M$ for
distinguishable particles, while for indistinguishable particles
$P_{\rm dble}^{\rm ind}=2/(M+1)$. Evidently $P_{\rm dble}^{\rm
ind}>P_{\rm dble}^{\rm dst}$, that is indistinguishable particles
are more likely to occupy the same state than distinguishable
particles.
\end{thebibliography}
\end{document}